\documentclass{article}
\usepackage[dvips]{graphicx} 
\usepackage{float,bm,physics,slashed}
\usepackage{amsmath,amssymb,tabularx,booktabs}
\usepackage{extpfeil}
\usepackage[titletoc]{appendix}
\usepackage[usestackEOL]{stackengine}
\usepackage{stackengine}
\usepackage{lipsum}
\edef\tmp{\the\baselineskip}
\setstackgap{L}{\tmp}
\usepackage{extarrows} 
\usepackage{color}
\usepackage[caption=false]{subfig}
\usepackage[colorlinks=true,linkcolor= blue,filecolor= blue,urlcolor= blue,citecolor= blue]{hyperref}
\usepackage{geometry} 
\usepackage{array,appendix}    
\begin{document}
\title{Sea quark-gluon effect on the magnetic octupole deformation of decuplet baryons}
\author{Preeti Bhall$^1$, Ritu Garg$^{*2}$ and Alka Upadhyay$^1$ \\ \small{\it $^{1}$Department of Physics and Material Science, Thapar Institute of Engineering and Technology, Patiala} \\ \small{\it $^{2}$Department of Physics, Manipal University Jaipur, Jaipur} \\ \small{Email: preetibhall@gmail.com, ritu.garg@jaipur.manipal.edu}}

\date{\today}

 \maketitle
\begin{abstract}
The magnetic octupole moment of  $J^P= \frac{3}{2}^+$ decuplet baryons are discussed in the statistical framework, treating baryons as an ensembles of quark-gluon Fock states. The probabilities associated with multiple strange and non-strange Fock states depict the importance of sea in spin, flavor $\&$ color space, which are further merged into statistical parameters. The individual contribution of valence and sea (scalar, vector and tensor) to the magnetic octupole moment is calculated. The symmetry breaking in both sea and valence is experienced by a suppression factor $k(1-C_l)^{n-1}$ and a mass correction parameter 'r', respectively. The factor $k(1-C_l)^{n-1}$ systematically reduces the probabilities of Fock states containing multiple strange quark pairs. The octupole moment value is obtained -ve for $\Delta^{++}, \Delta^+, \Sigma^{*+}$ and +ve for $\Delta^{-}, \Sigma^{*-}, \Xi^{*-}, \Omega^-$ baryons with the domination of scalar (spin-0) sea. The computed results are compared with existing theoretical predictions, demonstrating good consistency. These predictions may serve as valuable inputs for future high-precision experiments and theoretical explorations in hadron structure. 
\end{abstract}
\section{Introduction}
Understanding the electromagnetic profile of hadrons in terms of their constituent quarks, gluons, and sea quarks, remains a chronic challenge in particle physics. To comprehend the spatial and dynamic features of hadronic matter, electromagnetic Form Factors (FFs) serve as essential tools for probing the internal structure and characterizing their EM properties such as masses, charge radii, semi-leptonic decays quadrupole moment, and other high-order multipole moments.
A wealth of literature is dedicated to the study of static properties of octet ($J^P=\frac{1}{2}^+$) members using various theoretical approaches \cite{1,2,3,4,5,6,7,8,9,10,11,12}, comparatively less is known about the spin-$(\frac{3}{2})^+$ particles. On the experimental front, the magnetic moments of 7 octet baryons are measured with high-precision ($\sim 1\%$), whereas the $\Sigma^0 \rightarrow \Lambda$ transition magnetic moment is known with $\sim 5\%$ accuracy \cite{13, 14}. The charge radii of $J^P =\frac{1}{2}^+$ particles reported in the Particle Data Group (PDG) \cite{15} with values such as $r_p$ = 0.8409 $\pm$ 0.0004 fm,  $r^{2}_n$ = -0.115 $\pm$ 0.0017 fm$^2$, and $r_{\Sigma^{-}}$= 0.78 $\pm$ 0.10 fm. Furthermore, the quadrupole transition moment $(\Delta^+\rightarrow N)$ was extracted by LEGS \cite{16} and Mainz \cite{17} collaborations that provide the evidence of intrinsic deformities in the structure of nucleon and $\Delta^+$ baryon. The structural expolration of $\Delta$ and other decuplet members have been studied theoretically to a certain extent  \cite{18,19,20,21,22,23,24}, but the experimental data is very limited due to their short lifetimes. Only the magnetic moments of $\Delta^{++}$  and $\Delta^+$
have been determined through $\pi^+ p\rightarrow \pi^+ p\gamma$ \cite{25} and 
$\gamma p \rightarrow \pi^0 p \gamma'$ \cite{26} processes, with large error bars. Recent breakthrough came from the  BESIII \cite{27} and CLEO \cite{28}, which studied $\Omega^-$ baryon through the  $e^+e^- \rightarrow \Omega^-\bar\Omega^+$ reaction and characterizing the $\gamma^* \Omega^- \bar{\Omega}^+$ vertex via four Form Factors: electric charge ($|G_{E0}|$), magnetic dipole ($|G_{M1}|$), electric quadrupole ($|G_{E2}|$), and magnetic octupole ($|G_{M3}|$). In the coming years, remarkable advancements are expected in the decuplet sector. At MAMI \cite{27,29, 30}, and Jefferson Lab (JLab) \cite{31}, researchers plan to investigate the magnetic moment of the $N^*(1535)$ resonance, while JLab future program includes the EM measurements of $\Sigma^*$ and $\Xi^*$ hyperons \cite{32, 33}. The BESIII collaboration and a Super $J/\psi$ Factory (SCTF) are also anticipated to generate a secondary beam of $\Omega^-$ via $\psi(2S) \rightarrow \Omega^- \bar{\Omega}^+$ process \cite{34}.\\
Over the decades, majority of attention has been given to the study of lower-order EM moments such as magnetic moment, charge radii, electric quadrupole moment, and transition moments. However, the knowledge of higher-order moments, particularly the magnetic octupole moment, is scarce. The magnetic octupole moment (MOM) provides crucial findings about the spatial distribution of magnetization and reveals subtle aspects of charge and current configurations within composite particles.
Since no experimental studies exists for MOM, therefore a theoretical review of the literature is essential. In Refs. \cite{35,36}, the magnetic octupole moment of $\Delta$ baryons and hyperons was studied using the light cone QCD sum rules (LCSR) approach. Depending on the sign and magnitude, the current distributions are suggested to be of either prolate or oblate type. Using the  spectator quark model \cite{37}, G. Ramalho  predicted the value of electric quadrupole moment ($-0.043
efm^2$) and magnetic octupole moment ($-0.0035
efm^3$) of $\Delta^+$ and $\Omega^{-}$ \cite{38,39} baryon. Using the  In Refs. \cite{40,41}, A.J. Buchmann estimated EM multipole moments (charge radii, electric quadrupole moment, transition moments, magnetic octupole moment etc.) with the help of different models. Despite the analyses offered by theoretical models, the role of non-valence contributions - "sea quarks" remains largely unexplored in the context of higher-order electromagnetic moments. The sea is theorized to be a dynamic medium, comprising infinite $\bar q q$ pairs connected non-perturbatively to gluons. Recently, the STAR experiment at the Relativistic Heavy Ion Collider (RHIC), found the key role of sea antiquarks to the proton's spin \cite{42}. Also, the NuTeV collaboration \cite{43} at Fermilab not only confirmed the presence of strange quarks but also quantified their contribution to the nucleon spin. The measurement of ratio $\frac{2(s+ \bar s)}{u+ \bar u+ d+ \bar d}$ = 0.477 $\pm$ 0.063 $\pm$ 0.053 provided direct evidence for strange quark involvement.\\
The present work examined the role of sea quark in determining the magnetic octupole moment (MOM) of $J^P=\frac{3}{2}^+$ decuplet baryons. We employed a statistical framework in conjunction with the detailed balance principle, which has proven successful in computing the various static properties like masses \cite{44}, magnetic moment \cite{45}, charge radii \cite{46}, semi-leptonic decays \cite{47,48}, quadrupole moment and their transitions \cite{49,50}. In this approach, baryons are modeled as superpositions of quark-gluon Fock states. The detailed balance principle governs the probabilities of each Fock state within the hadron. Our primary goal is to analyze the contribution of strange and non-strange Fock states to the octupole moment. To experience the SU(3) symmetry breaking (valence and sea), a suppression factor $k(1-C_l)^{n-1}$ and a mass correction parameter are introduced that specifically probe strangeness effects. The factor $(1-C_l)^{n-1}$ limits the free energy of gluons which in turn affects the probabilities of Fock states. The individual contribution of sea components in terms of scalar, vector and tensor is also represented.\\
The outline of the present study is as follows: Sec. 2 provides a brief overview of the magnetic octupole moment. In Sec. 3, the decuplet baryon wavefunctions including sea components are discussed. Sec. 4 outlines the statistical model along with the principle of detailed balance. Sec. 5 discusses the numerical results of magnetic octupole moments. The conclusive remarks and future perspectives are summarized in Sec. 6.
\section{Magnetic Octupole moment}
The study of electromagnetic (EM) interactions is a fundamental and non-invasive approach for probing the spatial and dynamic characteristics of quarks and gluons in hadronic systems. These interactions are described through elastic and inelastic multipole Form Factors (FF's) \cite{51}. They contain useful information about the geometrical arrangements within hadrons, including their shape, size, spatial charge, and current distributions. The concept of shape is often discussed in analogy with nuclear physics, where deformed nuclei exhibiting rotational bands. In particular, spin- $\frac{1}{2}$ nuclei with vibrational and rotational modes exhibit vanishing static quadrupole moments in their ground states, yet their shape can be inferred from their excitation spectra. In case of hadrons like proton and $\Delta$, the situation is totally different as they are composed of quarks, with constant fluctuations of virtual $q\bar{q}$ pairs and meson clouds. 
As we know, most of the information about the $\Delta$ comes indirectly, from the study of the $\gamma N \rightarrow \Delta$ transition. The $\Delta$ elastic form factors are dominated by the electric charge ($G_{E0}$) and magnetic dipole ($G_{M1}$), while the electric quadrupole ($G_{E2}$) and magnetic octupole ($G_{M3}$) have relatively less contribution. In $\gamma N \Delta$ transition, the non-zero value of the $E_2$ and $C_2$ multipoles indicates the non-sphericity in the charge distribution inside the nucleon \cite{52,53}. The finite value of the quadrupole moment, along with axial and reflection symmetry, indicates a charge distribution shaped like a rugby ball (prolate) or a discus (oblate). Furthermore, the EM transitions are governed by the rule of angular momentum and parity conservation. The no. of multipole Form factors is limited by the conservation of angular momentum as:\\
$$|J_i-J_f|\leq J\leq J_i+J_f$$\\
Here, $J_i$ and $J_f$ represent the angular momentum of the initial and final states, respectively. In Table 1, the allowed elastic and inelastic multipole moments for $N(939), \Delta(1232)$, and $N^*(1680)$ resonances are listed.
\begin{table*}[h]{\normalsize
 \renewcommand{\arraystretch}{1.5}
 \tabcolsep 4.5mm
    \centering
   \small{
     \begin{tabular}{cccc}\hline \hline 
        \textbf{Baryonic state}  & \textbf{Elastic Multipole moments} & \textbf{Transition Multipole moments} \\\hline  
    $N(939)$&$C_0 (Monopole), M_1 (Dipole)$&None\\
    \hline  
$\Delta(1232)$&$C_0 (Monopole), C_2 (Quadrupole)$&$C_2 (Quadrupole),$\\
&$M_1 (Dipole), M_3 (Octupole)$&$M_1 (Dipole), E_2 (Quadrupole)$\\
\hline  
$N^{*}(1680)$&$C_0 (Monopole), C_2 (Quadrupole), C_4 (Hexadecapole)$&$C_2 (Quadrupole)$\\
&$M_1 (Dipole), M_3 (Octupole), M_5 (Pentadecapole)$& $E_2 (Quadrupole), M_3 (Octupole)$\\
\hline 
\hline   
    \end{tabular}}
    \caption{represents the allowed elastic and non-elastic multipole moments of baryonic states}
    \label{tab:my_label}
}
\end{table*}
For better understanding, it is important to move beyond quadrupole deformations. The magnetic octupole moment (MOM) is a higher-order EM property that characterizes the third-order complex spatial distributions of magnetic current inside the baryons. Being a quantum mechanical system, baryons has many interactions through the constituents quarks $\&$ gluons leading to a well-defined spin and parity. Therefore, their EM interactions are effectively described through EM multipole operators. The MOM operator $\widehat{\Omega}_B$, 
expressed in units of $[\text{fm}^2 \mu_N]$ is given by \cite{54} 
\begin{equation}  
\widehat{O}_B = \frac{3}{8} \int d^3r \, (3z^2 - r^2) \, ( \mathbf{r} \times \mathbf{J}(\mathbf{r}) )_z ,  
\end{equation}  
where $\mathbf{J}(\mathbf{r})$ represents the spatial current density and $\mu_N$ is the nuclear magneton.  
Notably, replacing the magnetic moment density $(\mathbf{r} \times \mathbf{J}(\mathbf{r}))_z$  
with charge density $\rho(\mathbf{r})$, the expression will be similar to the charge quadrupole moment \cite{9}. The MOM magnitude follows $\widehat{O} \simeq r^2 \mu$, where $\mu$ is the magnetic moment and $r^2$ characterizes its spatial extent, related to the quadrupole moment of the system. Deviations from spherical symmetry in the magnetic moment distribution reveal higher-order shape asymmetries. More precisely, for $\widehat{O} >$ 0, a prolate (cigar-shaped) magnetic moment distribution,
while for $\widehat{O} <$ 0, an oblate (pancake-shaped) magnetic moment distribution is observed as shown in Fig. 1.
\begin{figure}
    \centering
    \includegraphics[width=0.9\linewidth]{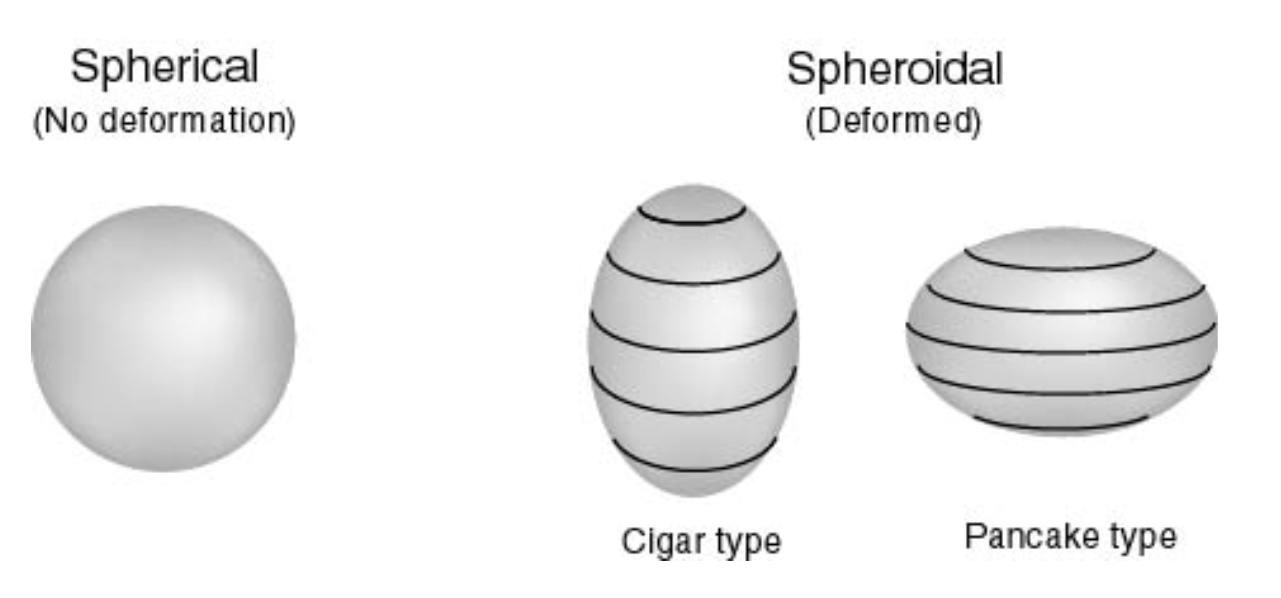}
    \caption{illustrates the charge distributions corresponding to positive (prolate- cigar shaped) and negative (oblate- pancake shaped) intrinsic quadrupole moment. A similar structural asymmetry is obtained for the octupole moment.
}
    \label{fig:enter-label}
\end{figure}
The magnetic octupole moment operator is given as:
\begin{equation}
 { \widehat{O}_B = C\sum_{i\neq j\neq k}^3 e_k(3\sigma_{iz}\sigma_{jz}-\sigma_i.\sigma_j)\sigma_k} 
\end{equation}
Here, C is a constant which characterize the matrix elements of color and orbital space. $\sigma_{iz}$ represent the z-component of Pauli spin matrices and $e_k$ is the charge of $k^{th}$ quark. The construction of this rank-3 tensor operator in spin space involves the incorporation of Pauli spin matrices of 3 distinct particles. 
The octupole moment operator ($\widehat{O}_B$) when applied to a specific wavefunction of baryons, they give particular coefficients to contribute to the octupole values. For this purpose, we must write the wavefunction of decuplet baryons in terms of quark-gluon Fock states leading to spin -$\frac{3}{2}$ with even parity, discussed in the next section. 

\section{Wavefunction for spin-$\frac{3}{2}^+$ decuplet baryons}
The non-relativistic quark model described the baryons as bound states of three valence quarks. The combination of three quarks ($q^3$) forms a color-neutral (singlet) state, in accordance with the principles of Quantum Chromodynamics (QCD). The valence quark wavefunction of the baryon \cite{55} is expressed as:     
\begin{equation}
    \Psi= \Phi (|\phi_{flavor}\rangle .|\chi_{spin}\rangle. |\psi_{color}\rangle.|\xi_{space}\rangle)
\end{equation}
The space-time $q^3$ wavefunction $|\xi\rangle$ is symmetric under the interchange of any two quarks. So, the flavor-spin-color must be antisymmetric under $q_i \leftrightarrow q_j$.
Beyond the traditional valence-only picture, it is assumed that a virtual 'sea' is also present, which plays a crucial role in shaping the internal configuration of baryons. Sea has a structure composed of an infinite no. of $q\bar q$ pairs with different flavors ($\bar u u, \bar d d, \bar s s$) and gluons. The 'sea' wavefunction has two components- spin (H) and color (G) that satisfy the condition: {$<H_i|H_j> = \delta_{ij}$, $<G_k|G_l> = \delta_{kl}$}. For simplicity, a flavorless sea is considered. If two gluons (each gluon spin-1) are assumed in sea, the possibilities in spin space has the expansion as:
$$\mathbf{Spin}: gg: 1 \bigotimes 1 = 0_s \bigoplus 1_a \bigoplus 2_s$$
In color space, each gluon with color octet ‘8’ will yield the following symmetric
and antisymmetric states:
$$\mathbf{Color}: gg: 8 \bigotimes 8 = 1_s \bigoplus 8_s \bigoplus 8_a \bigoplus {10}_a \bigoplus \bar{10}_a \bigoplus 27_s$$
Similar treatment is extended for multiple gluon and $\bar q q$ cases. Subscripts \textbf{\textit{a}} and \textbf{\textit{s}} denote the anti-symmetry and symmetry of the combined state. Since the baryon must be colorless and a $q^3$
state exist in color singlet (1), octet (8) and decuplet (10) states. This restricts the sea to be in a specific color state such that the combined wavefunction of baryon (valence + sea) remains a colorless entity. Therefore, the total wavefunction of decuplet baryons having spin-$\frac{3}{2}$, color singlet and flavor decuplet (10) represented as \cite{44}:\\
\begin{equation}
    \begin{aligned}
  |\Phi_{3/2}^{(\uparrow)}> =& \frac{1}{N} [a_0{\Phi_{1}}^{(\frac{3}{2} \uparrow)}H_{0}G_{1} + b_1({\Phi_1}^{(\frac{3}{2})} \otimes H_1)^{\uparrow} G_1 + b_8 ({\Phi_8}^{(\frac{1}{2})} \otimes H_1) ^{\uparrow} G_8 +\\&d_1 ({\Phi_1}^{(\frac{3}{2})} \otimes H_2)^{\uparrow} G_1 + d_8 ({\Phi_8}^{(\frac{1}{2})} \otimes H_2)^{\uparrow} G_8    
    \end{aligned}
\end{equation}
where $N^{2} = a_{0}^{2} +b_{1}^{2} +b_{8}^{2} + d_{1}^{2} + d_{8}^{2}$\\
Here N is the normalization constant. The first term of the wavefunction can be expanded as:\\
$$\Phi_{1}^{(\frac{3}{2} \uparrow)}= \Phi(10, \frac{3}{2},1)=F_S \psi_A$$\\
where
$$F_S=\phi^\lambda\chi^\lambda$$\\
$\Phi_{1}^{(\frac{3}{2} \uparrow)}$ is for the valence spin-$\frac{3}{2}$, color singlet state and 10 represent the flavor part. It can be written with the product of two wavefunctions $F_S$ and $\psi_A$ 
(contributing to the color of baryons
being anti-symmetric in nature). Further, $F_S$ denotes flavor ($\phi$) and spin ($\chi$) of the valence wavefunction. Following that, the sea spin ($H_0$) and color ($G_1$) wavefunctions are combined with the valence part. Similarly, the other terms $(\Phi_1^{(\frac{1}{2})} \otimes H_1)^\uparrow G_1$, $(\Phi_8^{(\frac{1}{2})} \otimes H_1)^\uparrow G_8$ etc. written with suitable C.G. coefficients by considering the symmetry property of the component wave function. The detailed explanation of the wavefunction can be found in Ref. \cite{44}. The parameters associated with each combination of the wavefunction $a_0, b_1, b_8, d_1, d_8$ are referred to as statistical coefficients. The coefficients have utmost importance as they encapsulate the contribution of different quark-gluon Fock states in spin, flavor and color space. Each coefficient reflects the probability amplitude of a particular sea configuration combining with the valence quark. Moreover, the sea quarks are classified based on their spin components, with spin-0, 1, and 2 corresponding to scalar, vector, and tensor sea, respectively.\\
Furthermore, the octupole operator (mentioned in eq. 2) depends on the spin and flavor of the quarks and is determined by calculating the matrix elements of the operator in the following way:
\begin{equation}
    \begin{aligned}
       \langle\Phi_{3/2}^{(\uparrow)}|{{\widehat{O}_B}}| \Phi_{3/2}^{(\uparrow)}\rangle =& \frac{1}{N^2} [{a_0^2}{\langle \Phi_{1}}^{(\frac{3}{2} \uparrow)}|\widehat{O}_B|{\Phi_{1}}^{(\frac{3}{2}\uparrow)}\rangle + {b_1^2} \langle{\Phi_1}^{(\frac{3}{2}\uparrow)}|\widehat{O}_B|{\Phi_1}^{(\frac{3}{2}\uparrow)}\rangle + \\& {b_8^2} \langle{\Phi_{8}}^{(\frac{1}{2}\uparrow)}|\widehat{O}_B|{\Phi_{8}}^{(\frac{1}{2}\uparrow)}\rangle + {d_1^2 }\langle{\Phi_1}^{(\frac{3}{2}\uparrow)}|\widehat{O}_B|{\Phi_1}^{(\frac{3}{2}\uparrow)}\rangle + + {d_8^2 }\langle{\Phi_8}^{(\frac{1}{2}\uparrow)}|\widehat{O}_B|{\Phi_8}^{(\frac{1}{2}\uparrow)}\rangle
    \end{aligned}
\end{equation}
The obtained expressions are given below:
\begin{subequations}
\begin{align}
\widehat{O}_{\Delta^{++}} &= a_{0}^2 (8C)+ b_{1}^2\left(\frac{56C}{15}\right)+d_{1}^2\left(\frac{-8C}{15}\right)+d_{8}^2\left(\frac{-28C}{5}\right) \\
\widehat{O}_{\Delta^{+}}  &= a_{0}^2 (4C)+ b_{1}^2\left(\frac{4C}{5}\right)+d_{1}^2\left(\frac{-12C}{5}\right)+d_{8}^2(-6C) \\
\widehat{O}_{\Delta^{-}}  &= a_{0}^2 (-4C)+ b_{1}^2\left(\frac{-12C}{15}\right)+d_{1}^2\left(\frac{12C}{5}\right)+d_{8}^2\left(\frac{14C}{5}\right)
\end{align}
\end{subequations}
\begin{subequations}
\begin{align}
\widehat{O}_{\Sigma^{*+}} &= a_{0}^2\left(\frac{C}{3}(16-4r)\right) + b_{1}^2\left(\frac{3C}{15}(16-4r)+\frac{2C}{15}(4r -16)\right)+ b_{8}^2\left(\frac{3C}{15}(16-4r)+\frac{3C}{15}(4r-16)\right) \notag \\ &\quad + d_{1}^2\left(\frac{C}{15}(16-4r)+\frac{4C}{15}(4r-16)\right) + d_{8}^2\left(\frac{C}{5}(-2r)+\frac{2C}{5}(8r)\right) \\
\widehat{O}_{\Sigma^{*-}} &= a_{0}^2\left(\frac{C}{3}(-8-4r)\right) + b_{1}^2\left(\frac{3C}{15}(-4r-8)+\frac{2C}{15}(8+4r)\right) + b_{8}^2\left(\frac{3C}{15}(-8-4r)+\frac{3C}{15}(8+4r)\right) \notag \\
&\quad + d_{1}^2\left(\frac{C}{15}(-8-4r)+\frac{4C}{15}(8+4r)\right) + d_{8}^2\left(\frac{C}{30}(-8-4r)+\frac{4C}{15}(8+4r)\right) \\
\widehat{O}_{\Sigma^{*0}} &= a_{0}^2\left(\frac{C}{6}(4-4r)\right) + b_{1}^2\left(\frac{C}{10}(4-4r)+\frac{2C}{15}(4r-4)\right) + b_{8}^2\left(\frac{C}{10}(4-4r)+\frac{3C}{15}(-4+4r)\right) \notag \\
&\quad + d_{1}^2\left(\frac{C}{30}(4-4r)+\frac{4C}{15}(4r-4)\right) + d_{8}^2\left(\frac{C}{60}(4-4r)+\frac{8C}{15}(-4+4r)\right)
\end{align}
\end{subequations}
\begin{subequations}
\begin{align}
\widehat{O}_{\Xi^{*-}} &= a_{0}^2\left(\frac{C}{3}(-4 - 8r)\right) + b_{1}^2\left(\frac{C}{5}(-4 - 8r) + \frac{2C}{45}(12 + 24r)\right) + b_{8}^2\left(\frac{C}{5}(-4 - 8r) + \frac{3C}{15}(4 + 8r)\right) \notag \\
&\quad + d_{1}^2\left(\frac{C}{15}(-4 - 8r) + \frac{4C}{45}(12 + 24r)\right) + d_{8}^2\left(\frac{C}{30}(-4 - 8r) + \frac{8C}{75}(4 + 8r)\right) \\
\widehat{O}_{\Xi^{*0}} &= a_{0}^2\left(\frac{C}{3}(-8r + 8)\right) + b_{1}^2\left(\frac{3C}{15}(-8r + 8) + \frac{2C}{15}(8r - 8)\right)  + b_{8}^2\left(\frac{C}{5}(-8r + 8) + \frac{3C}{15}(8r - 8)\right) \notag \\
&\quad + d_{1}^2\left(\frac{C}{15}(-8r + 8) + \frac{4C}{15}(8r - 8)\right) + d_{8}^2\left(\frac{C}{30}(-8r + 8) + \frac{4C}{15}(8r - 8)\right)
\end{align}
\end{subequations}
\begin{equation}
\begin{aligned}
\widehat{O}_{\Omega^{-}} &= a_{0}^2\left(C(-4r)\right) + b_{1}^2\left(\frac{3C}{5}(-4r) + \frac{2C}{5}(4r)\right)  + d_{1}^2\left(\frac{C}{5}(-4r) + \frac{4C}{5}(4r)\right) + d_{8}^2\left(\frac{C}{5}(-2r) + \frac{2C}{5}(8r)\right)
\end{aligned}
\end{equation}
The linear equations (6a)–(6c) represent the octupole moments of the $\Delta$ baryons in terms of the constant C and the statistical coefficients, with the exception of $\Delta^0$. In the case of  $\Delta^0$, all terms cancel out and yield a zero value for its octupole moment. Eqs. 7(a)-7(c), 8(a), 8(b) $\&$ 9 include an additional parameter 'r' because of strange quark. This parameter enables the study of flavor symmetry-breaking effects in the valence sector. In next section, a statistical method with the detailed balance principle is discussed to compute the probability distribution and thereby determine the statistical coefficients $(a_0, b_1, b_8, d_1, d_8)$.
\section{Statistical Model with Detailed balance principle}
The principle of detailed balance based on pure statistical considerations was introduced by Zhang et al. \cite{56}. It postulates the expansion of baryons in multiple quark-gluon Fock states. Each Fock state consists a continuum of $q\bar q$ pairs, mediated by gluons can be expressed as: \\
\begin{equation}
|h \rangle = \sum_{i,j,k,l} C_{i,j,k,l} | \{q^3\},\{i,j,k,l\}\rangle
\end{equation}
Here $\{q^3\}$ corresponds to the valence quarks, counting $i$, $j$, $l$ as the no. of  $u\bar{u}$, $d\bar{d}$, $s \bar s$ pairs respectively and $k$ represents the no. of gluons. The quarks and gluons within the Fock states are the intrinsic partons which are non-perturbatively connected to the gluons. The intrinsic partons are fundamentally differ from extrinsic partons that emerge transiently during high-energy scattering processes such as QCD hard bremsstrahlung and gluon splitting \cite{57}.\\
The probability of observing a baryon in a specific quark-gluon Fock state $|\{q^3)\},\{i,j,k,l\}\rangle$ is -
$$ \rho_{i,j,k,l} = |C_{i,j,k,l}|^2 $$
These probabilities $\rho_{i,j,k,l}$ must obey the normalization condition:\\
$$\sum_{i,j,k,l} \rho_{i,j,k,l} = 1$$
This principle asserts that the rate of transition into a given substate be equal to the rate of transition out of that substate.
$$\rho_{i,j,k,l} |\{q^{3}\},\{i,j,k,l\}\rangle \xrightleftharpoons{balance} \rho_{i',j',k',l'}|\{q^{3}\},\{i',j',k',l'\}\rangle$$
Different transition sub-processes like g $\rightleftharpoons  q \bar q$, g $\rightleftharpoons $ gg, and q $\rightleftharpoons $ qg are used to calculate the probabilities of various Fock states. In addition, the probabilities of Fock states are modified due to strange sea condensates. Importantly, the transition g $\rightleftharpoons  q \bar q$ is only permitted when the gluon energy ($\varepsilon_g$) exceeds twice the strange quark mass, i.e. $\varepsilon_g > 2M_s$. This energy threshold mandates reduced $\bar s s$ pair formation via gluon splitting. For this, a suppression factor is introduced, given by $k(1 - C_{l})^{n-1}$ \cite{58}, where n represents the total no. of partons present in the Fock state and $l$ denotes the no. of $\bar ss$ pairs. The value of parameter $C_{l-1}=\frac{2M_s}{M_B-2(l-1)M_s}$, where $M_B$ is the mass of baryon. The fascinating aspect of considering the strange sea is that it allows for the exploration of SU(3) flavor symmetry breaking within the sea.\\
For instance, with zero $\bar s s$ condensates in sea, the transition process proceeds as follows:
    \[
|\{q \},i,j,0,k\rangle
 \xrightleftharpoons[{l(l+1)}]{k(1-C_{0})^{n-2l-1}} |\{q \},i,j,1,k-1\rangle
\]\\
\[
\frac{\rho_{i,j,1,k-1}}{\rho_{i,j,0,k}} = \frac{k(1-C_0)^{n-2l-1}}{l(l+1)}
\]
And the sequence of transition process persists until all the gluons have been converted into strange $\bar qq$ pairs.
\[
|\{q \},i,j,k-1,1\rangle
 \xrightleftharpoons[{k(k+1)}]{1(1-C_{k-1})^{n-k-2}} |\{q \},i,j,k,0\rangle
\]\\
\[
\frac{\rho_{i,j,k,0}}{\rho_{i,j,k-1,1}} = \frac{1(1-C_{k-1})^{n-k-2}}{k(k+1)}
\]\\
The generalized expression for k no. of gluons follows as:
\begin{equation}
\frac{\rho_{i,j,l,k}}{\rho_{i,j,l+k,0}}= \frac{k(k-1)(k-2)...1(1-C_0)^{n-2l-1}(1-C_1)^{n-2l}....(1-C_{l-1})^{n+k-2}}{(l+1)(l+2)...(l+k)(l+k+1)}
\end{equation}
Using the normalization condition, the individual probabilities of the Fock states is determined in terms of \(\rho_{0000}\) \cite{45,47}. The importance of factor $(1- C_l)^{n-1}$ becomes more apparent when sea strange $\bar q q$ condensates are considered in connection with singly and doubly strange baryons \cite{44,45},  thereby impacting the probability distribution of corresponding Fock states. The Fock states without strange quark content make up 86$\%$ of the total Fock states, while the inclusion of $s\bar s$ reduces this proportion to 80$\%$. It is important to note, although sea admits infinitely many Fock state configurations, practical calculations require limiting the quark-gluon content. This is because states with higher complexity make small contributions to baryon observables.\\
Thereafter, the statistical formalism is applied to evaluate the relative probabilities of Fock states in spin and color space. The wave function mentioned in eq. (4) can be reformulated as \textbf{$\Phi_{\text{val}} \Phi_{\text{sea}}$},
and the statistical parameters  $(a_0, b_1, b_8, d_1, d_8)$ are scaled by a factor $(\sum  n^{'}_{\mu \nu} c^{\text{sea}})$ and the complete wave function is expressed as:
\[
|\Phi^{\uparrow}_{3/2} \rangle = \sum_{\mu, \nu} \left( n^{'}_{\mu \nu} c^{\text{sea}}\right) \Phi_{\text{val}} \Phi_{\text{sea}},
\]
where \(\mu = 0, 1, 2\) representing spin states and \(\nu = 1, 8, 10\) denoting color states. All \(n^{'}_{\mu \nu}\) values are computed based on the multiplicities of each Fock state and expressed in the form of relative probability \(\rho_{p,q}\). Here, $p$ corresponds to the core quark spin while $q$ represents the sea quark spin, constrained such that their combination yields a total spin of 3/2 and maintains an overall color singlet state. Each statistical coefficient carries a specific value of \(\sum n_{\mu\nu} \, c^{\text{sea}}\) depending on the Fock states (e.g. $|u \bar ug\rangle$, $|d \bar dg\rangle$, $|s \bar sg\rangle$, $|u \bar u d\bar d \bar ss \rangle$, $|s \bar s d\bar d gg\rangle$).
The common parameter 'c' is derived from the flavor probabilities. For comprehensive details of these calculations, we refer readers to Refs. \cite{44,57}. By integrating all the probabilities, the statistical coefficients ($a_{0}, b_{1},b_{8}, d_1, d_8$) are determined.
These coefficients capture the influence of sea quarks to the various static properties of octet and decuplet baryons.
\section{Numerical analysis and discussion}
In this section, we present the calculated magnetic octupole moments (MOMs) for spin-$\frac{3}{2}^+$ baryons belonging to the decuplet family (such as $\Delta, \Sigma^{*}, \Xi^{*}$, and $\Omega^-$ particles). The results are summarized in Table 2. The non-zero values of octupole moment suggest a deviation from spherical symmtery in the current distribution of baryons. The analysis adopts a statistical framework in which the octupole moment is formulated as a set of linear equations comprising statistical coefficients ($a_0, b_1, b_8, d_1, d_8$) and a fitting parameter C. The optimal value of C is obtained through $\chi^2$ minimization method, ensuring the best possible agreement with existing theoretical predictions. Sea quark dynamics are embedded within the statistical coefficients. The independent effect of sea components i.e. scalar (spin-0), vector (spin-1) and tensor (spin- 2) sea is presented in Table 3. Each spin state is corresponds to a specific set of statistical coefficients. In order to obtained the separate contribution, selectively suppression is applied:
$$\textbf{Scalar sea} (a_0 \neq 0, b_1, b_8, d_1, d_8=0 )$$
$$\textbf{Vector sea} (b_1, b_8 \neq 0, a_0, d_1, d_8=0 )$$
$$\textbf{Tensor sea} (d_1,d_8 \neq 0, b_1, b_8, a_0 =0 )$$ The statistical model accounts for the probabilistic nature of quark-gluon Fock state and is analyzed in three configurations i.e. Model C, P $\&$ D. Model C is the basic model and assumed the equal probability of each
Fock state with definite spin and color quantum numbers. Model D and P are the modified versions of Model C. Model D exhibits a systematic suppression mechanism for higher-multiplicity Fock state contributions within the color-spin configuration space. It means that the more crowded the Fock states, the less likely they survive due to higher interactions.  Model P, in contrast, incorporates a restriction where $\bar q q$ pair can
form colorless pseudo-scalar Goldstone bosons. These internal bosons introduce additional symmetry to the Fock states and obviates the active role of sea quark within the baryons. We investigated the magnetic octupole moments for spin-$\frac{3}{2}^+$ decuplet particles using Model C and D.  
\par\vspace{\baselineskip}\noindent
\textbf{(i)} It is clearly observed from Table 3 that for all the decuplet members, the scalar sea acts as the only active contributor from the total sea. This behavior may be due to the higher multiplicities of valence spin states. When the spin of core quarks coupled with the sea spin (i.e., spin-0, 1, and 2), the formation of spin- $\frac{3}{2}$ is more favored with spin-0 (scalar sea) as compared to spin - 1 (vector) and spin- 2 (tensor). The scalar sea (spin-0) contributes over $90\%$ of the MOM. While the effect of tensor sea (spin-2) is minimal, it can be considered for few particles. The vector sea (spin-1), however, has no significant effect. Similarly, in the case of Model C, the scalar sea emerges as the sole dominant contributor among the sea components. Both vector and tensor sea components make no contribution. \\
For $\Delta^{++}$ baryon, the highest value of MOM is observed, reflecting the contribution of both valence and sea. The neutral particles ($\Delta^0, \Sigma^{*0}, \Xi^{*0}$) have zero value, consistent with charge symmetry expectations. It is interesting to put forward that smaller value of octupole moment is observed for doubly strange ($\Xi^{*-} =uss, \Xi^{*0}=dss$) particles as compared to singly strange ($\Sigma^{*+}= uus, \Sigma^{*-} = dds$) baryons as shown in Table 2. This could be possible due to the suppression factor $k(1-C_l)^{n-1}$, which restricts the available free energy of gluons. The accommodation of large no. of $\bar s s$ condensates within higher mass baryons leads to a reduction in this factor. As a result, it lowers the overall probability of that particular Fock state and change the value of octupole moment.
\par\vspace{\baselineskip}\noindent
\textbf{(ii)} The Standard Model describe the flavor SU(3) symmetry breaking in the terms of current quark masses i.e. $m_u, m_d << m_s$ . A mass correction parameter 'r' is proposed that directly includes the strange quark mass to the relevant operator and is defined as r = $\frac{\mu_s}{\mu_d}$ \cite{55}. $\mu_s$ and $\mu_d$ are the magnetic moments of the strange and down quark, respectively. The value of parameter 'r' has been calculated (r=0.850) in previous work \cite{49,50}. Notice that, initially, $\Sigma^{*0}, \Xi^{*0}$ baryons yield zero quark contribution. And a non-vanishing contribution is observed to the octupole moments of neutral decuplet baryons after considering breaking effect. It indicates that asymmetry in quark masses affect the internal dynamics and magnetic distributions. Relative to the symmetry expectations, the value of magnetic octupole moment deviates by around $4–10\%$. The SU(3) limit highlights a few consistent relations observed in both Model C and D.
\begin{equation}
   \frac{1}{2} \widehat{O}^{\text{SU(3)}}_{\Delta^{++}} = \widehat{O}^{\text{SU(3)}}_{\Delta^{+}} = \widehat{O}^{\text{SU(3)}}_{\Delta^{-}} = \widehat{O}^{\text{SU(3)}}_{\Sigma^{*+}} = -\widehat{O}^{\text{SU(3)}}_{\Sigma^{*-}} = -\widehat{O}^{\text{SU(3)}}_{\Xi^{*-}} = - \widehat{O}^{\text{SU(3)}}_{\Omega^{-}} \end{equation}
   \begin{equation}
\widehat{O}^{\text{SU(3)}}_{\Sigma^{*0}}=\widehat{O}^{\text{SU(3)}}_{\Xi^{*0}}= \widehat{O}^{\text{SU(3)}}_{\Delta^{0}}
    \end{equation}
Additional expression that are valid (upto three digits) in the exact SU(3) symmetry given below:
\begin{equation}
\widehat{O}_{\Delta^{++}} +\widehat{O}_{\Delta^{+}}- \widehat{O}_{\Delta^{-}}+\widehat{O}_{\Delta^{0}}=0
\end{equation}
\begin{equation}
\widehat{O}_{\Delta^{+}} +\widehat{O}_{\Delta^{-}}=0
\end{equation}
\begin{equation}
\widehat{O}_{\Sigma^{*+}}+ \widehat{O}_{\Sigma^{*-}} =0    
\end{equation}
\begin{equation}
   \widehat{O}_{\Xi^{*}} + \widehat{O}_{\Omega^{-}} + \widehat{O}_{\Delta^{++}} =0 
\end{equation}
After symmteric breaking, the following pattern is observed:
\begin{equation}
\widehat{O}_{\Delta^{++}} > \widehat{O}_{\Sigma^{*+}} > -\widehat{O}_{\Delta^{-}}> \widehat{O}_{\Delta^{+}} >-\widehat{O}_{\Sigma^{*-}} > -\widehat{O}_{\Xi^{*-}} >-\widehat{O}_{\Omega^{-}}>  \widehat{O}_{\Xi^{*0}}> \widehat{O}_{\Sigma^{*0}}
\end{equation}
Moreover, both Model C and D yield results that differ by approximately $2–8\%$,  accentuate a slight but meaningful difference in their predictions.
\begin{table*}[h!]{\normalsize
\renewcommand{\arraystretch}{1.0}
\tabcolsep 0.1cm
    \centering
   \small{
   \begin{tabular}{ccccccc}\toprule\hline
           &           & \multicolumn{2}{c}{Model D} &  &\\\cmidrule{3-4}
   Particles & \Longunderstack{Octupole\\ moment}& \Longunderstack{Valence\\+ \\Sea ($g\rightarrow u\bar u, d\bar d, s\bar s$)}& \Longunderstack {Symmetry breaking \\ in valence \\(r=0.850)}& \Longunderstack{Model C\\ with sea+valence\\(r=0.850)}& \Longunderstack{Without\\sea} \\\midrule
   \toprule \vspace{2mm}
    $\Delta^{++}$  & 7.6492 C &-0.0249&-0.0249&-0.0260&-0.0182\\ \vspace{2mm}
    $\Delta^{+}$  & 3.6522 C &-0.0119&-0.0119&-0.0130&0.0011\\ \vspace{2mm}
    $\Delta^{0}$  &-&0.0&0.0&0.0&0.0\\ \vspace{2mm}
    $\Delta^{-}$  & -3.7258 C &0.0121&0.0121&0.0130&-0.0013\\ \vspace{2mm}
    $\Sigma^{*+}$  &C (4.9005 - 1.2251 r)&-0.0119&-0.0125&-0.0136&0.0013\\ \vspace{2mm}
    $\Sigma^{*-}$ &C (-2.4502 - 1.2251 r)&0.0119&0.0113&0.0123&-0.0012\\ \vspace{2mm}
    $\Sigma^{*0}$ &C (0.5957 - 0.5957 r)&0.0&-0.0002&-0.0003&0.0014\\ \vspace{2mm}
    $\Xi^{*-}$ &C (-1.2301 - 2.4603 r) &0.0120&0.0108&0.0117&0.0044\\ \vspace{2mm}
    $\Xi^{*0}$ &C (2.4596 - 2.4596 r)&0.0&-0.0012&-0.0013&0.0001\\ \vspace{2mm}
    $\Omega^{-}$ &C (-3.8838 r)&0.0126&0.0107&0.0110&-0.0011\\ \hline \bottomrule              \end{tabular}
              }}
         \caption{Octupole moment of spin-$\frac{3}{2}^+$ decuplet baryons in the units of [fm$^3$] with the value of C$'$=-0.0032}
         \label{tab:my_label}
     \end{table*}

     \begin{table*}[h!]{\normalsize
 \renewcommand{\arraystretch}{1.5}
 \tabcolsep 4.5mm
    \centering
   \small{
     \begin{tabular}{ccccccccc}\toprule \hline 
        \textbf{Particles} &\textbf{Scalar sea} &\textbf{Vector sea} &\textbf{Tensor sea}   &\textbf{Model D}&\textbf{Model C}\\
        & \textbf{(spin-0)}&\textbf{(spin-1)}&\textbf{(spin-2)}&\textbf{(scalar+tensor)}&\textbf{(scalar)}\\\hline  
$\Delta^{++}$ &-0.02494&-0.00009&0.00006&-0.02488&-0.02607\\\hline
$\Delta^{+}$ &-0.01225&-0.00003&0.00038&-0.01187&-0.01303\\\hline 
$\Delta^{0}$ &0.0&0.0&0.0&0.0&0.0\\\hline 
$\Delta^{-}$ &0.01243&0.00002&-0.00030&0.01213&0.01303\\\hline 
 $\Sigma^{*+}$& -0.01295&-0.00001&0.00039&-0.01256&-0.01368\\
\hline  
$\Sigma^{*-}$& 0.01172&0.00001&-0.00035&0.01137&0.01238\\
 \hline  
$\Sigma^{*0}$& -0.00031&0.0&0.00001&-0.0003&-0.00032\\
\hline 
 $\Xi^{*-}$&0.01110&0.00002&-0.00030&0.0108&0.01170\\
\hline 
$\Xi^{*0}$&-0.00123&0.0&0.00003&-0.0012&-0.00130\\
\hline 
$\Omega^{-}$ &0.01087&0.0&-0.00011&0.01076&0.01108\\\hline\bottomrule 
 \end{tabular}}
    \caption{Individual contribution of sea components}
    \label{tab:my_label}
}
\end{table*}

 \begin{table*}[h!]{\normalsize
 \renewcommand{\arraystretch}{1.5}
 \tabcolsep 2.5mm
    \centering
   \small{
     \begin{tabular}{c|c|c|c|c} \toprule\hline 
        Particles  &  Model D& Model C&   GPM\cite{40}& QCD Sum rule\cite{35}\\ \hline  
    $\Delta^{++}$&-0.0249&-0.0260&-0.024&-0.006$\pm$0.002 \\      
        $\Delta^{+}$ &-0.0119 &0.0117&-0.012&-0.003$\pm$0.001\\  
 $\Delta^{0}$&0.0&0.0&0.0&0.0\\  
 $\Delta^{-}$&0.0121&0.0130&0.012&0.003$\pm$0.001\\
$\Sigma^{*+}$&-0.0125&-0.0136&-0.004&-0.015$\pm$0.005\\ 
$\Sigma^{*-}$&0.0113&0.0123&0.008&0.013$\pm$0.004\\ 
$\Sigma^{*0}$&-0.0002&-0.0003&0.002&-0.001$\pm$0.0003\\ 
$\Xi^{*-}$&0.0108&0.0117&0.005&0.020$\pm$0.006\\ 
$\Xi^{*0}$&-0.0012&-0.0013&0.002&-0.0014$\pm$0.0005\\ 
$\Omega^{-}$&0.0107&0.0110&0.003&0.016$\pm$0.004\\ \hline\bottomrule
 \end{tabular}}
    \caption{Comparison of our computed results with different theoretical approaches}
    \label{tab:my_label}
}
\end{table*} 
\par\vspace{\baselineskip}\noindent
\textbf{(iii)} The statistical formalism predicted the -ve value of MOM for positively charged baryons ($\Delta^{++}, \Delta^+, \Sigma^{*+}$) and a +ve value for the negatively charged ones ($\Delta^-, \Sigma^{*-}, \Xi^-, \Omega^-$). This pattern suggests the prolate and oblate current distribution for $\Delta^-, \Sigma^{*-}, \Xi^-, \Omega^-$ and $\Delta^{++}, \Delta^+, \Sigma^{*+}$ respectively. For instance, if the sea is completely excluded from the statistical approach, the results deviate by more than $80\%$. This highlights the crucial role of the sea in shaping the spatial charge and current configurations within baryons. Since no experimental data currently exists for the MOM of spin-$\frac{3}{2}^+$ decuplet baryons. Therefore, we compared our results with the available theoretical predictions presented in Table 4. The results show a good consistency in sign and magnitude with the Refs. \cite{35,40}. 
\section{Conclusion}
The present study focuses on the prediction of magnetic octupole moment (MOM) of decuplet ($\frac{3}{2}^+$) baryons using statistical techniques. The dynamics of sea are captured through the probabilities of various Fock states, represented by statistical coefficients ($a_0, b_{1}, b_{8}, d_{1}, d_{8}$). We obtained non-zero magnetic cotupole moments, demonstrating that their current distribution deviates from spherical symmtery. The $\Delta^{++}, \Delta^+, \Sigma^{*+}$ exhibits -ve value while the $\Delta^{-}, \Sigma^{*-}, \Sigma^{*0}, \Xi^{*-}, \Xi^{*0}, \Omega^-$ members show the +ve value of MOM, which correspond to the oblate and prolate current distributions respectively. Unlike lower-order moments, octupole moments are sensitive to more complex deformations and asymmetries in the hadron's structure. The flavor symmetry breaking in sea and valence is also studied. An inverse correlation is noted between the strangeness and magnetic octupole moment due to the suppression factor $k(1-C_l)^{n-1}$. The individual effects of scalar, vector, and tensor sea are analyzed, revealing the dominance of scalar sea (spin-0) on the magnetic octupole moment. Notably, the exclusion of sea entirely from the model deviates the computed results up to 80$\%$.  A key advantage of this statistical framework is that it requires no additional parameters. The calculations are performed in a non-relativistic regime at 1 GeV$^2$ energy scale. Interestingly, octupole deformation has been observed in atomic nuclei, leading to "pear-shaped" structures in distinct nuclei \cite{59}. This observation raises the intriguing possibility that similar deformation effects could be explored in particle physics in the future.


\begin{thebibliography}{63}

\bibitem{1}
Panda, A. R. and Roy, K. C. and Sahoo, R. K., \href{https://link.aps.org/doi/10.1103/PhysRevD.49.4659}{Phys. Rev. D} \textbf{49}, 9 (1994)
\bibitem{2}
N. Barik and M. Das, \href{https://doi.org/10.1016/0370-2693(83)90475-6}{Physics Letters B} \textbf{120},4 (1983)
\bibitem{3}
Zhang, Jun and Ma, Bo-Qiang, \href{https://link.aps.org/doi/10.1103/PhysRevC.93.065209}{Phys. Rev. C} \textbf{93}, 6 (2016)
\bibitem{4}
G. Kälbermann and J.M. Eisenberg, \href{https://doi.org/10.1016/0370-2693(90)90882-7}{Physics Letters B} \textbf{247}, 12 (1990)
\bibitem{5}
A.J. Buchmann, \emph{\href{https://www.worldscientific.com/doi/abs/10.1142/9789812776914_0021}{Phenomenology of Large $N_c$ QCD}}, pp. 224--228 (2002)
\bibitem{6}
Kaur, Amanpreet and Gupta, Pallavi and Upadhyay, Alka, \href{https://doi.org/10.1093/ptep/ptx068}{PTEP} \textbf{2017}, 6 (2017)
\bibitem{7}
X. Y. Liu, K. Khosonthongkee, A. Limphirat, Y. Yan, \href{https://doi.org/10.1088/0954-3899/41/5/055008}{J. Phys. G} \textbf{41}, 055008 (2014)
\bibitem{8}
Singh, Harpreet and Kumar, Arvind and Dahiya, Harleen, \href{https://dx.doi.org/10.1088/1674-1137/41/9/094104}{Chinese Physics C} \textbf{41}, 9 (2017)
\bibitem{9}
A.J. Buchmann and E.M. Henly
\href{https://doi.org/10.1103/PhysRevC.63.015202}{Phys. Rev. C} \textbf{63}, 015202 (2000)
\bibitem{10}
M. Batra and A. Upadhyay., \href{https://doi.org/10.1016/j.nuclphysa.2013.11.008}{Nuclear Physics A} \textbf{922}, pp. 126-139 (2013)
\bibitem{11}
N.~Sharma, H.~Dahiya, \href{https://doi.org/10.1007/s12043-012-0479-y}{Pramana} \textbf{80}, 237 (2013)
\bibitem{12}
H.~Dahiya, \href{https://dx.doi.org/10.1088/1674-1137/42/9/093102}{Chinese Physics C} \textbf{42}, 093102 (2018)
\bibitem{13}
Contreras, J.~G., Huerta, R. and Quintero, L.~R., {Revista Mexicana de Fisica} \textbf{50}, 5 (2004)
\bibitem{14}
 Petersen, P. C. et al., \href{https://link.aps.org/doi/10.1103/PhysRevLett.57.949}{Phys. Rev. Lett.} \textbf{57}, 8 (1986)
 \bibitem{15}
 Navas, S. and others, \href{10.1103/PhysRevD.110.030001}{Phys. Rev. D} \textbf{110}, 3 (2024)
 \bibitem{16}
G.~Blanpied, M.~Blecher, A.~Caracappa et al. (The LEGS Collaboration), \href{https://doi.org/10.1103/PhysRevLett.79.4337} {Phys. Rev. Lett.} \textbf{79}, 4337 (1997).
 \bibitem{17}
 L.~Tiator, D.~Drechsel, S.S. Kamalov and S.N. Yang, \href{https://doi.org/10.1140/epja/i2002-10177-6} {Eur. Phys. J.A} \textbf{17}, 3 (2003).
 \bibitem{18}
 Sahoo, R. K. and Panda, A. R. and Nath, A., \href{https://link.aps.org/doi/10.1103/PhysRevD.52.4099}{Phys. Rev. D.} \textbf{52}, 7 (1995)
 \bibitem{19}
 A.J. Buchmann, E.M. Henley, \href {https://link.aps.org/doi/10.1103/PhysRevD.65.073017} {Phys. Rev. D} \textbf{65}, 073017 (2002)
 \bibitem{20}
 J.Y. Kim, H.C. Kim, G.S. Yang, M.~Oka, \href{https://link.aps.org/doi/10.1103/PhysRevD.103.074025}{Phys. Rev. D} \textbf{103}, 074025 (2021)
 \bibitem{21}
 R.H. Abada.A, Weigel~H, \href{https://doi.org/10.1016/0370-2693(95)01353-9}{Physics Letters B}  (1996)
 \bibitem{22}
 Y.~OH, \href{ https://doi.org/10.1142/S0217732395001137}{Modern Physics Letters A} \textbf{10}, 1027 (1995)
 \bibitem{23}
 M.N. Butler, M.J. Savage, R.P. Springer, \href{https://link.aps.org/doi/10.1103/PhysRevD.49.3459}{Phys. Rev. D} \textbf{49}, 3459 (1994)
 \bibitem{24}
 Hacker, C., Wies, N., Gegelia, J. et al.,
\href{https://doi.org/10.1140/epja/i2006-10043-7}{Eur. Phys. J. A} \textbf{28}, 5-9 (2006) 
\bibitem{25}
G. López Castro and A. Mariano, \href{https://doi.org/10.1016/S0370-2693(01)00980-7}{Physics Letters B} \textbf{517}, 3 (2001)
\bibitem{26}
M. Kotulla et al., \href{https://link.aps.org/doi/10.1103/PhysRevLett.89.272001}{Phys. Rev. Lett.} \textbf{89}, 27 (2002)
\bibitem{27}
M. Ablikim et al. (BESIII Collaboration), \href{https://link.aps.org/doi/10.1103/PhysRevD.107.052003}{Phys. Rev. D} \textbf{107}, 052003 (2023)
\bibitem{28}
S. Dobbs and A. Tomaradze and T. Xiao and Kamal K. Seth and G. Bonvicini,  \href{https://doi.org/10.1016/j.physletb.2014.10.025}{Phys. Lett. B} \textbf{739}, pp. 90-94 (2014)
\bibitem{29}
T.M. Aliev, M. Savci, \href{ https://doi.org/10.1103/PhysRevD.90.116006}{Phys. Rev. D} \textbf{90}, 116006 (2014)
\bibitem{30}
B. Krusche and S. Schadmand, \href{https://doi.org/10.1016/S0146-6410(03)90005-6}{Progress in Particle and Nuclear Physics} \textbf{51}, pp. 399-485 (2003)
\bibitem{31}
V. Punjabi et. al (Jefferson Lab Hall A Collaboration), \href{https://link.aps.org/doi/10.1103/PhysRevC.71.055202}{Phys. Rev. C} \textbf{71}, 055202 (2005)
\bibitem{32}
L. Guo, D. P. Weygand, M. Battaglieri, R. De Vita, V. Kubarovsky, P. Stoler, M. J. Amaryan, P. Ambrozewicz, M. Anghinolfi et al. (CLAS Collaboration), \href{https://doi.org/10.1103/PhysRevC.76.025208}{Phys. Rev. C} \textbf{76}, 025208 (2007)
\bibitem{33}
J.T. Goetz et al., \href{https://doi.org/10.1103/PhysRevC.98.062201}{Phys. Rev. C} \textbf{98}, 062201 (2018)
\bibitem{34}
C.Z. Yuan, M. Larliner, \href{https://doi.org/10.1103/PhysRevLett.127.012003}{Phys. Rev. Lett.} \textbf{127}, 012003 (2021)
\bibitem{35}
T.M. Aliev and K. Azizi and M. Savcı, \href{https://doi.org/10.1016/j.physletb.2009.10.026}{Physics Letters B} \textbf{681}, 3 (2009)
\bibitem{36}
K. Azizi, \href{https://doi.org/10.1140/epjc/s10052-009-0988-0}{Eur. Phys. J. C } \textbf{61}, 311–319 (2009) 
\bibitem{37}
G. Ramalho and M.T. Peña and Franz Gross, \href{https://doi.org/10.1016/j.physletb.2009.06.052}{Physics Letters B} \textbf{678}, 4 (2009)
\bibitem{38}
Ramalho, G., \href{https://link.aps.org/doi/10.1103/PhysRevD.103.074018}{Phys. Rev. D} \textbf{103}, 074018 (2021)
\bibitem{39}
Ramalho, G. and Pe\~na, M. T., \href{https://link.aps.org/doi/10.1103/PhysRevD.83.054011}{Phys. Rev. D} \textbf{83}, 054011 (2011)
\bibitem{40}
A.J. Buchmann, \href{https://doi.org/10.1007/s00601-018-1464-x}{Few-Body Systems} \textbf{59}, 145 (2018)
\bibitem{41}
A.J. Buchmann, E.M. Henley, \href{https://doi.org/10.1007/978-3-540-85144-8_39}{Eur. Phys. J.A} \textbf{35}, pp. 267-269 (2008)
\bibitem{42}
Adam, J. et al., \href{https://link.aps.org/doi/10.1103/PhysRevD.99.051102}{Phys. Rev. D} \textbf{99}, 051102 (2019)
\bibitem{43}
 M.Goncharov et al., \href{https://link.aps.org/doi/10.1103/PhysRevD.64.112006}{Phys. Rev. D} \textbf{64}, 112006 (2001)
\bibitem{44}
A.~Kaur, A.~Upadhyay, \href{https://doi.org/10.1140/epja/i2016-16332-6} {The European Physical Journal A} \textbf{52}, 332 (2016)
\bibitem{45}
A.~Kaur, P.~Gupta, A.~Upadhyay, \href{https://doi.org/10.1093/ptep/ptx068}{Progress of Theoretical and Experimental Physics} \textbf{2017}, 063B02 (2017)
\bibitem{46}
P.~Bhall, A.~Upadhyay, \href{https://doi.org/10.1093/ptep/ptae060} {Progress of Theoretical and Experimental Physics} \textbf{2024}, 053B04 (2024)
\bibitem{47}
M.~Batra, A.~Upadhyay, \href{https://www.sciencedirect.com/science/article/pii/S0375947413007914}{Nuclear Physics A} \textbf{922}, 126 (2014)
\bibitem{48}
A.~Upadhyay, M.~Batra, \href {https://doi.org/10.1140/epja/i2013-13160-2} {The European Physical Journal A} \textbf{49}, 160 (2013)
\bibitem{49}
P.~Bhall, M.~Batra, A.~Upadhyay, \href{https://doi.org/10.1093/ptep/ptad108}{Progress of Theoretical and Experimental Physics} \textbf{2023}, 093B03 (2023)
\bibitem{50}
Bhall, P., Upadhyay, A., \href{https://doi.org/10.1140/epjc/s10052-025-13925-4}{Eur. Phys. J. C} \textbf{85}, 238 (2025)
\bibitem{51}
M. K. Jones, K. A. Aniol, F. T. Baker, J. Berthot, P. Y. Bertin, W. Bertozzi, et al. (The Jefferson Lab Hall A Collaboration), \href{https://doi.org/10.1103/PhysRevLett.84.1398}{Phys. Rev. Lett.} \textbf{84}, 1398 (2000)
\bibitem{52}
Bernstein, A. M. and Papanicolas, C. N., \href{}{AIP Conf. Proc} \textbf{904}, 1 (2007)
\bibitem{53}
V. Pascalutsa and M. Vanderhaeghen and S. N. Yang, \href{https://doi.org/10.1016/j.physrep.2006.09.006}{Physics Reports} \textbf{437}, 5 (2007)
\bibitem{54}
Donnelly, T William and Sick, Ingo, \href{https://doi.org/10.1103/RevModPhys.56.461}{Reviews of modern physics} \textbf{56}, 3 (1984)
\bibitem{55}
X.~Song, V.~Gupta, \href{https://link.aps.org/doi/10.1103/PhysRevD.49.2211} {Phys. Rev. D} \textbf{49}, 2211 (1994)
\bibitem{56}
Y.J. Zhang, W.Z. Deng, B.Q. Ma, \href {https://link.aps.org/doi/10.1103/PhysRevD.65.114005}{Phys. Rev. D} \textbf{65}, 114005 (2002)
\bibitem{58}
Y.J. Zhang, B.Q. Ma, L.M. Yang, \href{ https://doi.org/10.1142/S0217751X03014915} {International Journal of Modern Physics A} \textbf{18}, 1465 (2003)
\bibitem{57}
J.P. Singh, A.~Upadhyay, \href{https://dx.doi.org/10.1088/0954-3899/30/7/005}{Journal of Physics G: Nuclear and Particle Physics} \textbf{30}, 881 (2004)
\bibitem{59}
Butler, P.A., Gaffney, L.P., Spagnoletti, P. et al., \href{https://doi.org/10.1038/s41467-019-10494-5}{Nature Communications} \textbf{10}, 1 (2019)



\end{thebibliography}
\end{document}